\documentclass[showpacs,preprint]{revtex4}
\usepackage{graphicx}
\usepackage{subfigure}
\usepackage{amssymb}
\usepackage{amsmath}
\usepackage{bm}
\usepackage{latexsym}
\usepackage{hyperref}

\newcommand{\Tr}{{\text Tr}}
\begin{document}
\title{Entanglement evolution for excitons of two separate quantum dots in
a cavity driven by magnetic field}
\author{Jun Jing\footnote{Email: jingjun@sjtu.edu.cn}, Z. G. L\"{u}, H. R. Ma}
\affiliation{Institute of Theoretical Physics,
Shanghai Jiao Tong University\\
800 DongChuan Road, MinHang, Shanghai 200240, China}
\date{\today}

\begin{abstract}
The time evolution of entanglement for excitons in two quantum
dots embedded in a single mode cavity is studied in a
``spin-boson'' regime. It is found that although with the
dissipation from the boson mode, the excitons in the two quantum
dots can be entangled by only modulating their energy bias
$\epsilon$ under the influence of external driving magnetic field.
Initially, the two excitons are prepared in a pure separate state.
When the time-dependent magnetic field is switched on, a highly
entangled state is produced and maintained even in a very long
time interval. The mechanism may be used to control the quantum
devices in practical applications.

\end{abstract}
 \maketitle

\section{Introduction}

Since the age of \emph{EPR} paradox \cite{Einstein}, quantum
entanglement, as the magic non-local correlation of quantum system
revealed by the violation of Bell's inequality \cite{Bell}, had
attracted many attentions. During the past two decades, entangled
states provide great potential in the quantum communication and
information process \cite{Bennett, Nielsen}. It is also regarded as
the most intriguing and inherent feature of quantum composite system
and the resource of quantum computation \cite{Grover}. Thus the
preparation of entangled pair in a practical way and the maintenance
of its entanglement degree in a comparatively long time are
important issues for the realization of qubits and quantum gate.\\

There is significant interest in quantum information processing
using semiconductor quantum dot. Recently, there are many works
devoted to the quantum information properties, such as the
entanglement creation, of semiconductor quantum dot, since it is a
qualified qubit candidate \cite{Xia}. In comparison with other
physical systems \cite{Braunstein}, solid-state devices, in
particular, ultra-small quantum dots \cite{Burkard} with spin
degrees of freedom embedded in nanostructured materials are more
easily scaled up to large registers and they can be manipulated by
energy bias and tunneling potentials \cite{Loss}. Burkard et al.
\cite{Burkard} consider a quantum-gate mechanism based on electron
spins in coupled quantum dots, which provides a general source of
spin entanglement. Liu et al. \cite{Liu, Miranowicz} investigated
the generation of the maximally entanglement for coherent excitonic
states in two coupled large quantum dots (``large'' means $R\gg
a_B$, where $R$ is the radius of each dot and $a_B$ is Bohr radius
of excitons in quantum dots) mediated by a cavity field, which is
initially prepared in an odd coherent state. Zhang et al.
\cite{Zhang} showed that the Bell states and GHZ states can be
robustly generated by manipulating the system parameters. Chen et
al. \cite{Chen} observed that the photon trapping phenomenon in
double quantum dots (their distance is small compared to the emitted
photon $\lambda$) generates a entangled state. Nazir et al.
\cite{Nazir} found that two initially nonresonant quantum dots may
be brought into resonance by the application of a single detuned
laser. This allowed for the generation of highly entangled excitonic
states on the $10^{-12}$ second time scale. The possibility of
creating spin quantum entanglement in a system of two electrons
confined respectively in two vertically coupled quantum dots in the
presence of spin-orbit interaction had been explored by Zhao et al.
\cite{Zhao}. However, most of their successful works were devoted to
producing entanglement from two coupled quantum dots, but there are
few schemes to create a highly entangled state for excitons in
separate quantum dots from a pure separate state by only external
periodical field. \\

Cavity QED experiments, where few atoms are coupled to single cavity
modes, have culminated in the demonstration of creation of
entanglement between three distinguishable quantum systems
\cite{Raimond}. Indeed an integrated cavity QED consisted of two
quantum dots is more practical and realizable \cite{Badolato}. This
paper is applied for the generation of an entangled excitonic state
in the system of two identical quantum dots placed in a single-mode
cavity. The number of electrons excited from the valence-band to the
conduction-band in each dot is assumed to be small and the
exciton-exciton interaction in the same dot can be neglected
\cite{Wang}. Then the subsystem in which we hope to create
entanglement is consisted of two exciton picked from the two quantum
dots respectively. Each of them are simplified by considering only
the ground state and the first excited state of it. Then the whole
system is approximated like a spin-boson model. The distance of the
two dots is large enough to ignore the direct coupling between two
excitons and they are connected only through the cavity mode. The
measurement of the subsystem was chosen to be the concurrence found
by Wootters' group \cite{Wootters1, Wootters2} in the year of 1997
and von Neumann entropy. It will be demonstrated that in our model,
the entanglement of the two excitons can be enhanced to a high
degree by the external field. And we also try to clarify the physics
behind it. The rest of this paper is organized as follows. In
section \ref{Hamiltonian}, we introduce the Hamiltonian for our
``spin-boson'' model and describe the computation procedure for the
time-evolution of the concurrence and quantum entropy for the
subsystem; Detailed results and discussions are in section
\ref{discussion}; The conclusion of our study is given in section
\ref{conclusion}.

\section{The Hamiltonian and the theory}\label{Hamiltonian}

Essentially, the model we studied is two separate two-level atoms
(spins) connected by a single-mode boson gas under an periodical
magnetic field along $\vec{z}$ direction. We assume that the the two
``atom'' is identical and their interactions with the boson are
identical. The Hamiltonian of this model can be written as:
\begin{eqnarray}\label{Hami}
H &=& H_0+H_t, \\
H_0 & = &
\sum_{i=1}^2(-\frac{\Delta_i}{2}\sigma^x_i+\frac{\epsilon_i}{2}\sigma^z_i)+
\omega(a^+a+\frac{1}{2})+g\sum_{i=1}^2(a+a^+)\sigma^x_i, \\
H_t & =& \sum_{i=1}^2F(t)\sigma^z_i,
\end{eqnarray}
where $\epsilon$ is the energy bias and $\Delta$ is the tunneling
potential. The time-dependent periodical (the period is $P$)
external field $F(t)$ could be a rectangular wave, a triangular
wave or a cosine wave $F(t)=A\cos(\frac{2\pi}{P}t)$, wherein $A$
is the amplitude, $P$ is the period length. $\omega$ is the
frequency of the single-mode and $\omega\ll\epsilon$. $g$ is the
spin-boson coupling strength and $g\ll1.0$. $\sigma_x$ and
$\sigma_z$ are the well-known Pauli matrix:
\begin{equation}
\sigma^x=\left(\begin{array}{cc}
      0 & 1 \\
      1 & 0
    \end{array}\right), \quad
\sigma^z=\left(\begin{array}{cc}
      1 & 0\\
      0 & -1
    \end{array}\right).
\end{equation}

The whole state of the whole system $\rho(t)$ can be formally
calculated by such a process:
\begin{eqnarray}
\rho(t)&=&\exp(-iHt)\rho(0)\exp(iHt),\\
\rho(0)&=&\rho_{12}(0)\otimes\rho_b(0),\\
\rho_{12}(0)&=&|\psi(0)\rangle\langle\psi(0)|,\\
\rho_b(0)&=&|\phi(0)\rangle\langle\phi(0)|,
\end{eqnarray}
where the initial state of the two spins $\psi(0)$ is a separate
state, for instance, $\psi(0)=|0\rangle_11\rangle_2$, both
$|0\rangle$ and $|1\rangle$ are the eigenstates in the space of
$\{\sigma^2, \sigma^z\}$; $\phi(0)$ is the initial state of the
boson mode, which is set as a vacuum state $|0\rangle$. The
evolution operator $U(t)=\exp(-iHt)$ can be evaluated by the
efficient algorithm of polynomial schemes \cite{Dobrovitski1, Hu,
Jing}. The method used here is the Laguerre polynomial expansion
method we proposed in Ref. \cite{Jing}, which is pretty well suited
to this problem and can give accurate result in a comparatively
smaller computation load. More precisely, the evolution operator
$\exp(-iHt)$ is expanded in terms of the Laguerre polynomial of the
Hamiltonian as:
\begin{equation*}
U(t) = e^{-iHt} = \left(\frac{1}{1+it}\right)^{\alpha+1}
\sum^{\infty}_{k=0}\left(\frac{it}{1+it}\right)^kL^{\alpha}_k(H),
\end{equation*}
where $\alpha$ distinguishes different types of Laguerre polynomials
\cite{Arfken}, $k$ is the order of the Laguerre polynomial. In real
calculations the expansion has to be cut at some value of
$k_{\text{max}}$, which was taken to be $20$ in this study. With the
largest order of the expansion fixed, the time step $t$ is
restricted to some value in order to get accurate results of the
evolution operator. For longer times the evolution can be achieved
by more steps. The action of the Laguerre polynomial of Hamiltonian
to the states is calculated by recurrence relations of the Laguerre
polynomial. The efficiency of it is about $8$ times faster than
conventional methods such as Runge-Kutta algorithm. \\

After obtaining the density matrix of the whole system, the reduced
density matrix of the two two-level excitons can be found by a
partial trace operation to $\rho(t)$, which traces out the degrees
of freedom of the single-mode boson:
\begin{equation}\label{final}
\rho_{12}(t)=\Tr_b(\rho(t)).
\end{equation}
For the model of this paper, $\rho_{12}(t)$ can be expressed as a
$4\times4$ matrix in the Hilbert space of the subsystem spanned by
the orthonormal vectors $|00\rangle$, $|01\rangle$, $|10\rangle$ and
$|11\rangle$. \\

The main tool of discussing pairwise entanglement is the concept of
concurrence. The definition of concurrence between spins 1 and 2 can
be found in \cite{Wootters1, Wootters2}:
\begin{equation}\label{Concurrence}
C=\max\{\lambda_1-\lambda_2-\lambda_3-\lambda_4,~0\},
\end{equation}
It is a monotone of von-Neumann entropy. $\lambda_i$, $i=1,2,3,4$
are the square roots of the eigenvalues of the product matrix
$\rho_{12}\tilde{\rho}_{12}$ in decreasing order. Where
$\tilde{\rho}_{12}$ is constructed as
$(\sigma_y\otimes\sigma_y)\rho^*(\sigma_y\otimes\sigma_y)$. Equation
(\ref{Concurrence}) applies for both of the mixed state and pure
state. For a maximally entangled state, $C=1$, and for a separate
one, $C=0$.

\section{Results and discussions}\label{discussion}

In this section, we showed the time evolutions of concurrence for
$\rho_{12}$ under different kinds of periodical magnetic fields in
comparison with that with no field. The energy bias and tunneling
potential in Hamiltonian (\ref{Hami}): $\epsilon=\Delta=0.4$. The
amplitude $A$ and the period $P$ of rectangular wave, triangular
wave and cosine wave are $A=0.48, P=4.0$. The other parameters are
kept as: $\omega=0.02$, $g=0.02$.\\

\begin{figure}[htbp]\centering
\subfigure[$|\psi(0)\rangle=|00\rangle$, rectangular field]{
\label{figent00:rec}
\includegraphics[width=3in]{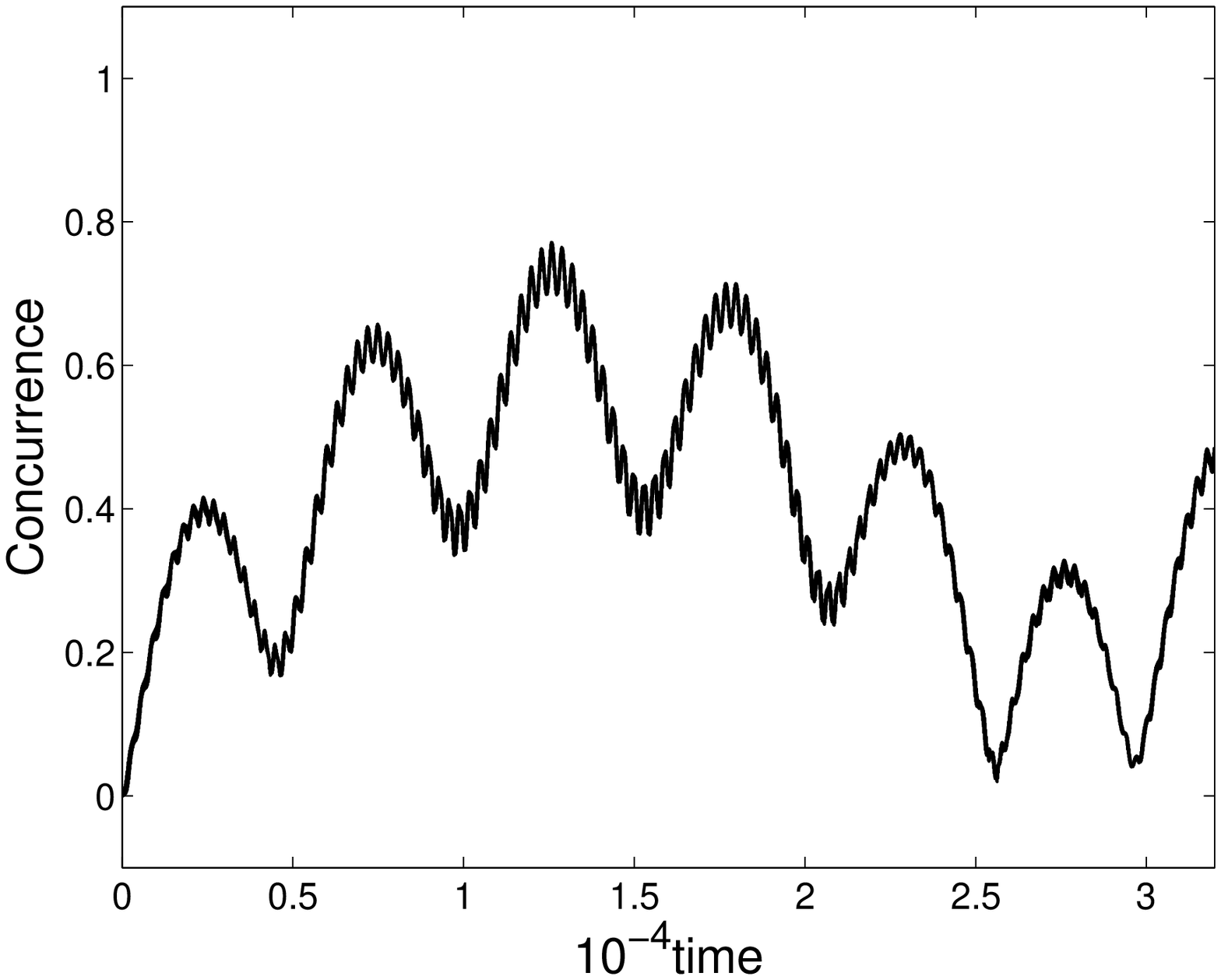}}
\subfigure[$|\psi(0)\rangle=|00\rangle$, cosine field]{
\label{figent00:cos}
\includegraphics[width=3in]{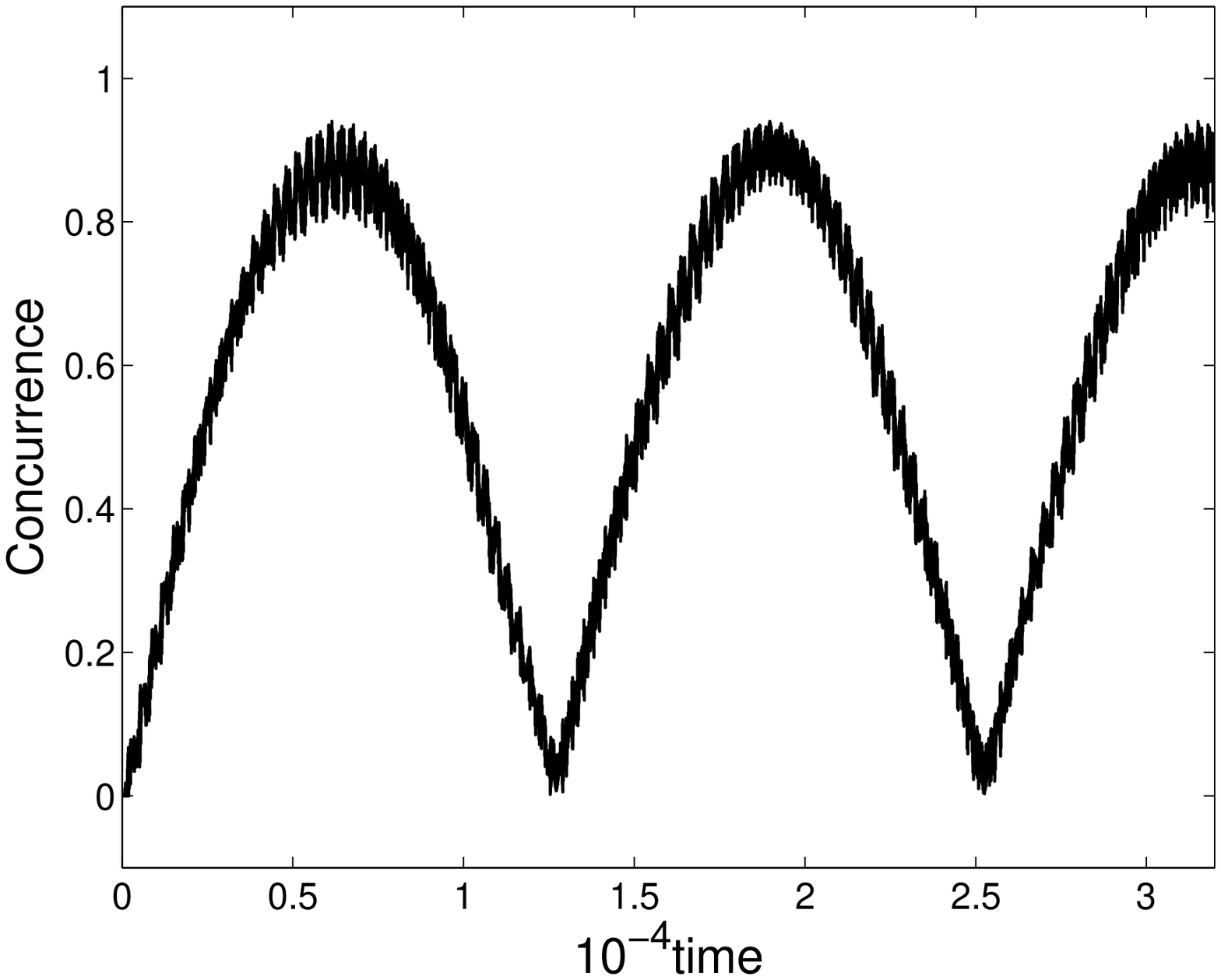}}
\subfigure[$|\psi(0)\rangle=|00\rangle$, triangular field]{
\label{figent00:tri}
\includegraphics[width=3in]{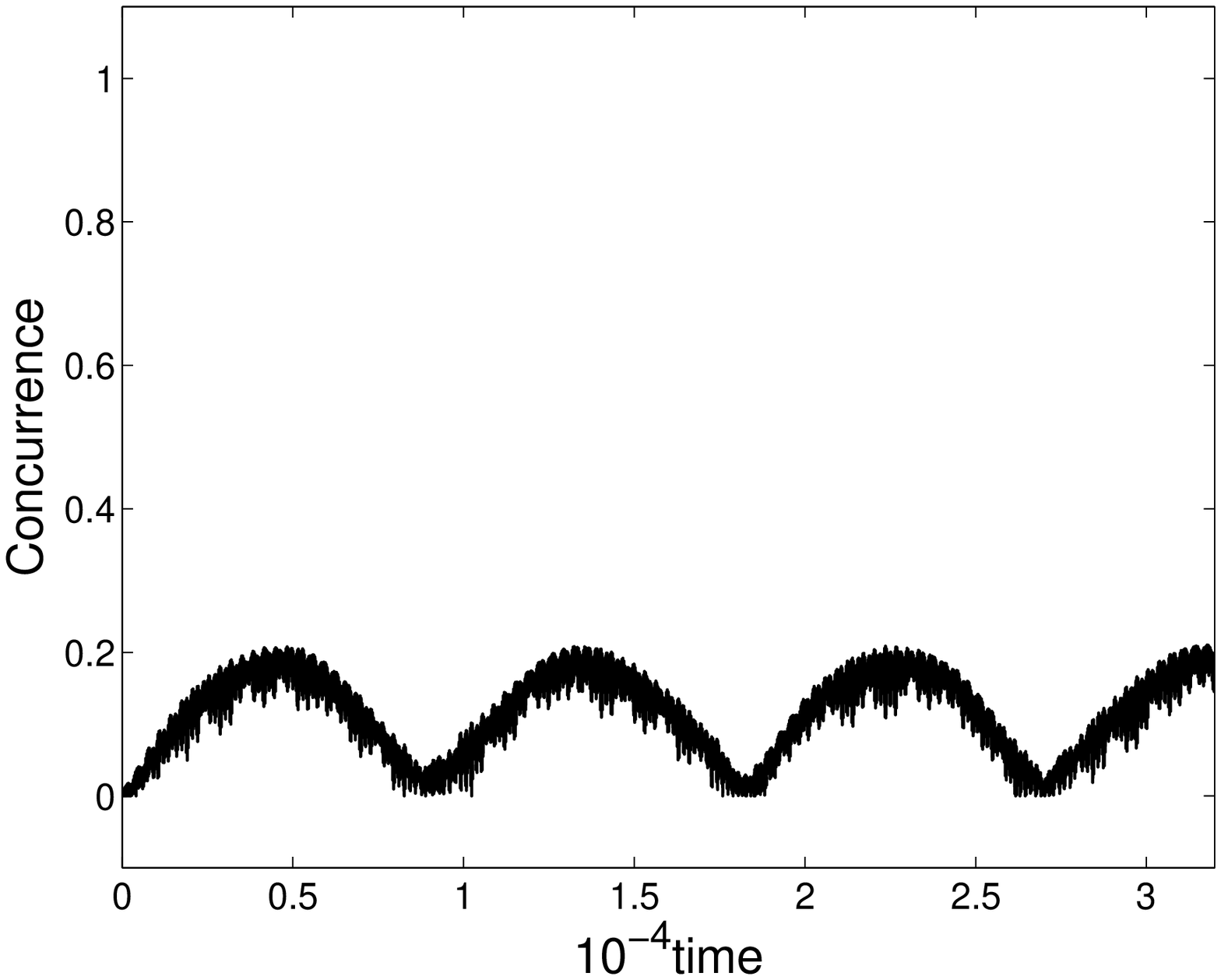}}
\subfigure[$|\psi(0)\rangle=|00\rangle$, no field $A=0$]{
\label{figent00:std}
\includegraphics[width=3in]{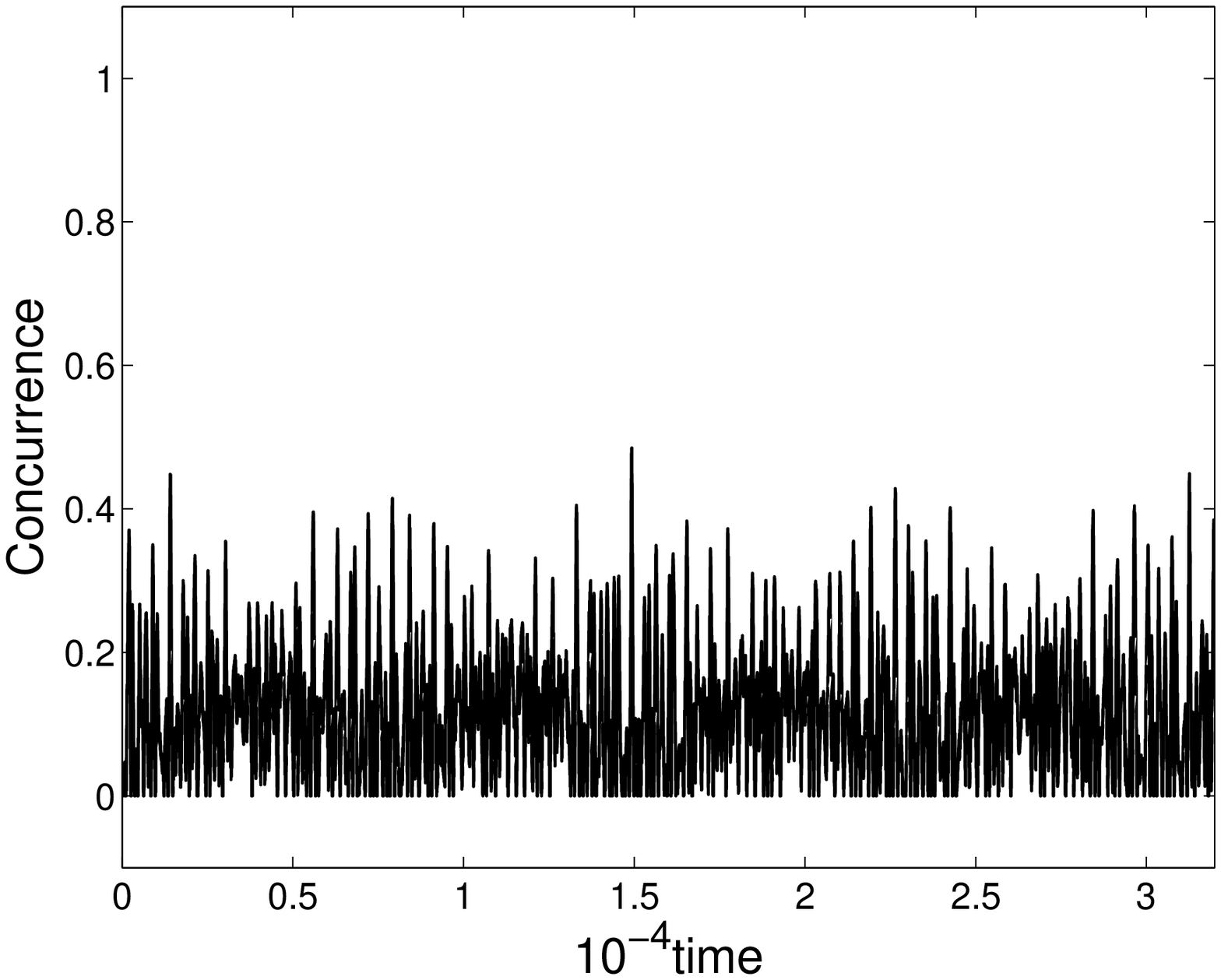}}
\caption{The time-evolution of entanglement for
$|\psi(0)\rangle=|00\rangle$ under different field vs. the pure
dissipation process without any modulating action} \label{figent00}
\end{figure}

\begin{figure}[htbp]\centering
\subfigure[$|\psi(0)\rangle=|01\rangle$, rectangular field]{
\label{figent01:rec}
\includegraphics[width=3in]{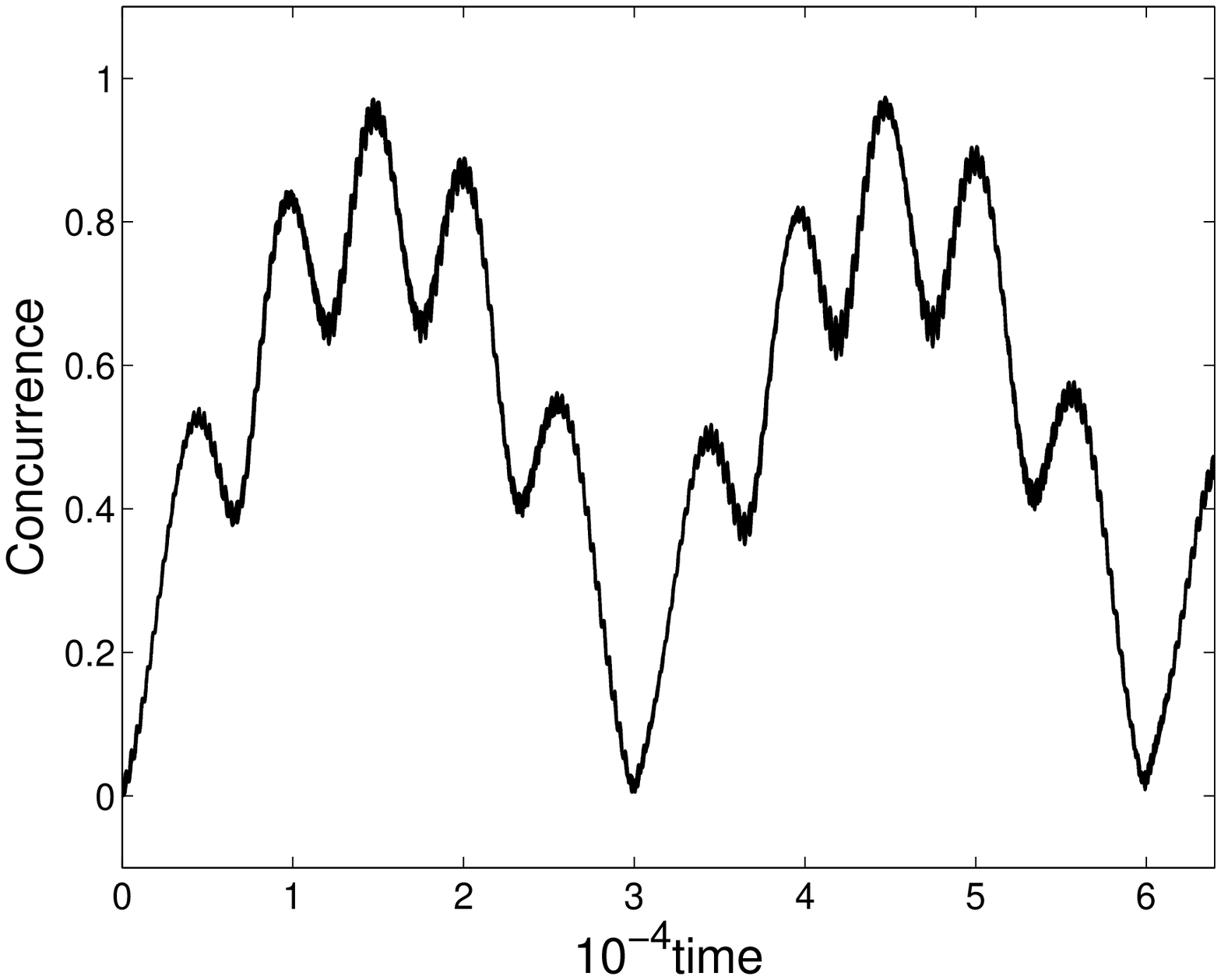}}
\subfigure[$|\psi(0)\rangle=|01\rangle$, cosine field]{
\label{figent01:cos}
\includegraphics[width=3in]{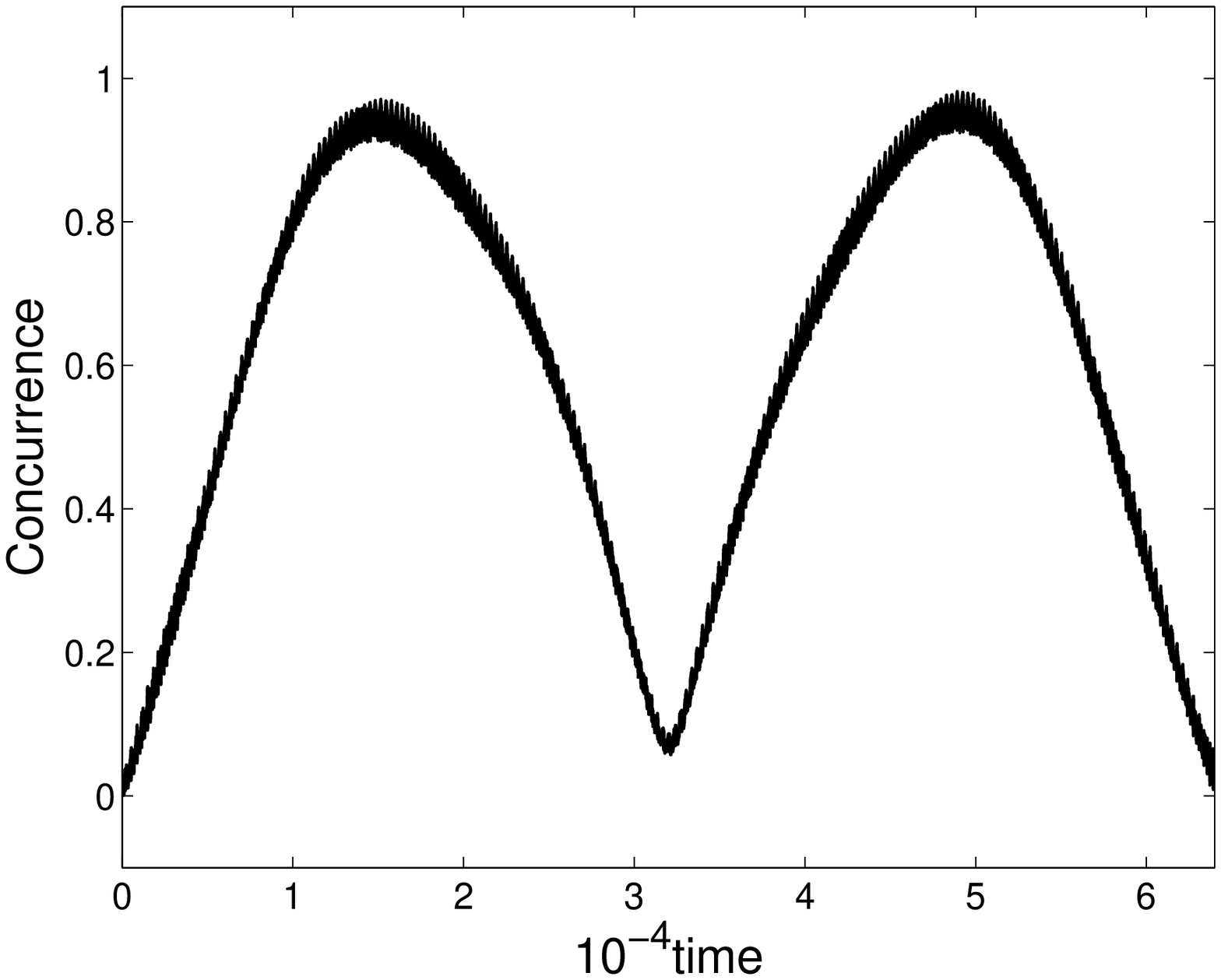}}
\subfigure[$|\psi(0)\rangle=|01\rangle$, triangular field]{
\label{figent01:tri}
\includegraphics[width=3in]{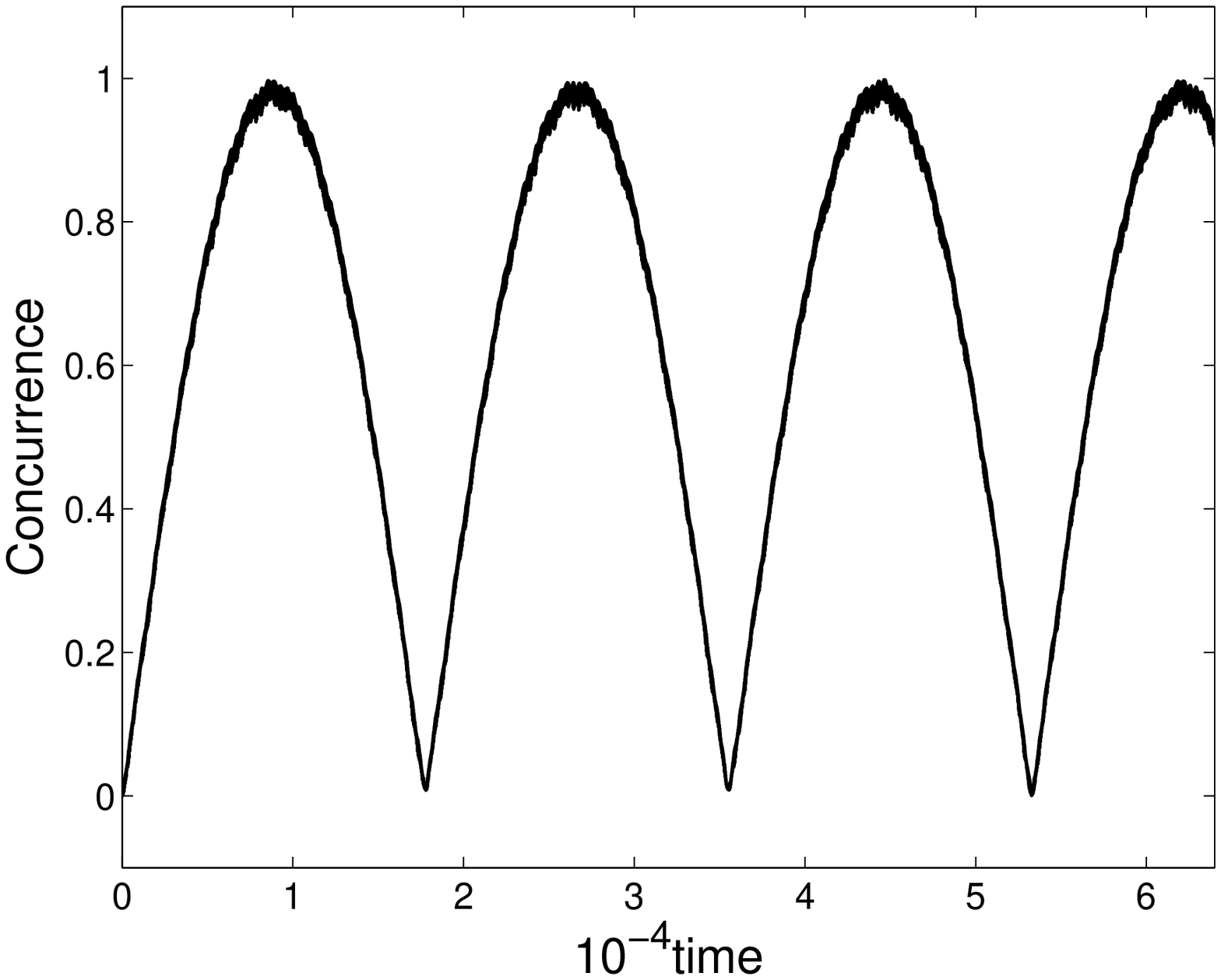}}
\subfigure[$|\psi(0)\rangle=|01\rangle$, no field $A=0$]{
\label{figent01:std}
\includegraphics[width=3in]{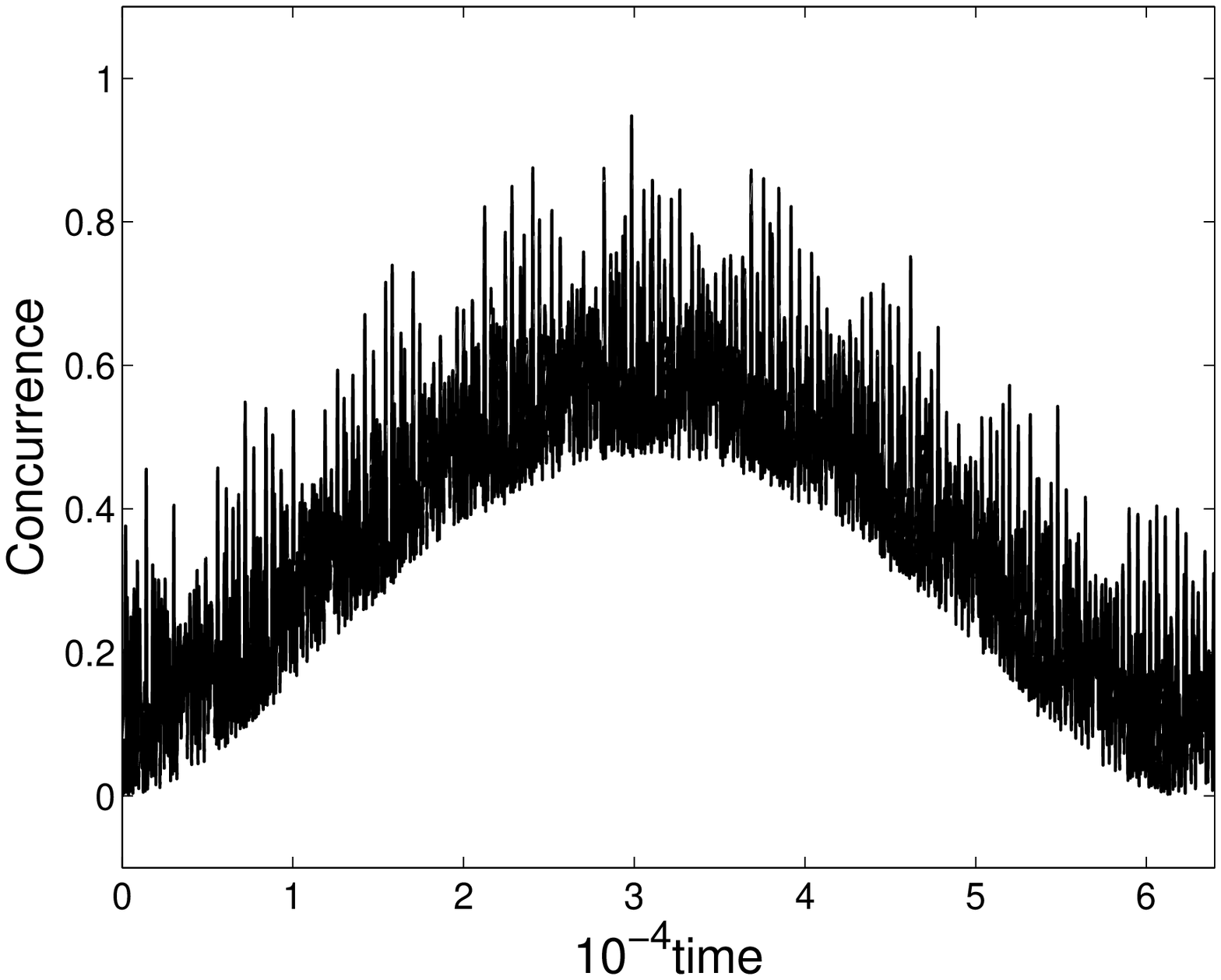}}
\caption{The time-evolution of entanglement for
$|\psi(0)\rangle=|01\rangle$ under different field vs. the pure
dissipation process without any modulating action}\label{figent01}
\end{figure}

\begin{figure}[htbp]\centering
\subfigure[$|\psi(0)\rangle=|11\rangle$, rectangular field]{
\label{figent11:rec}
\includegraphics[width=3in]{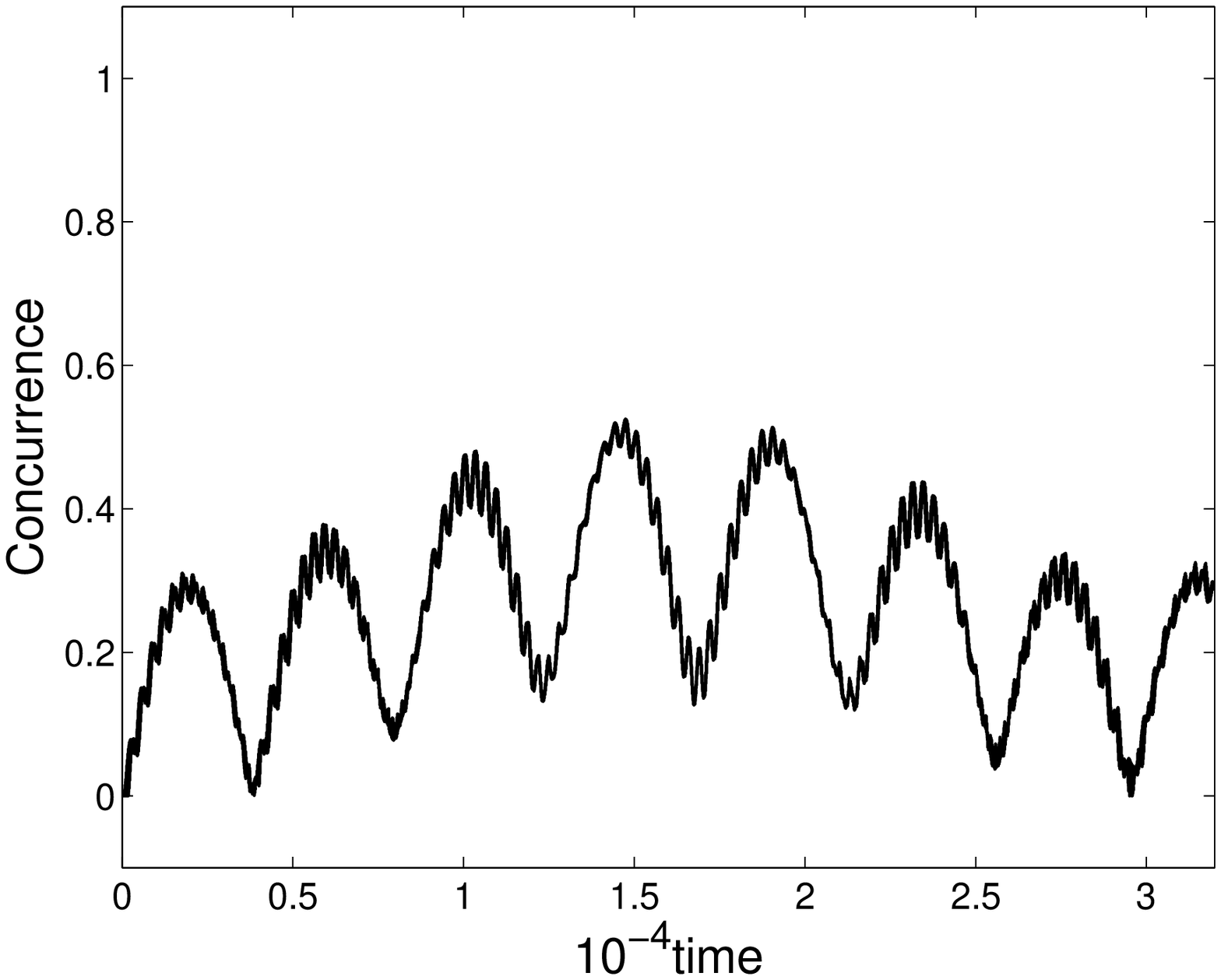}}
\subfigure[$|\psi(0)\rangle=|11\rangle$, cosine field]{
\label{figent11:cos}
\includegraphics[width=3in]{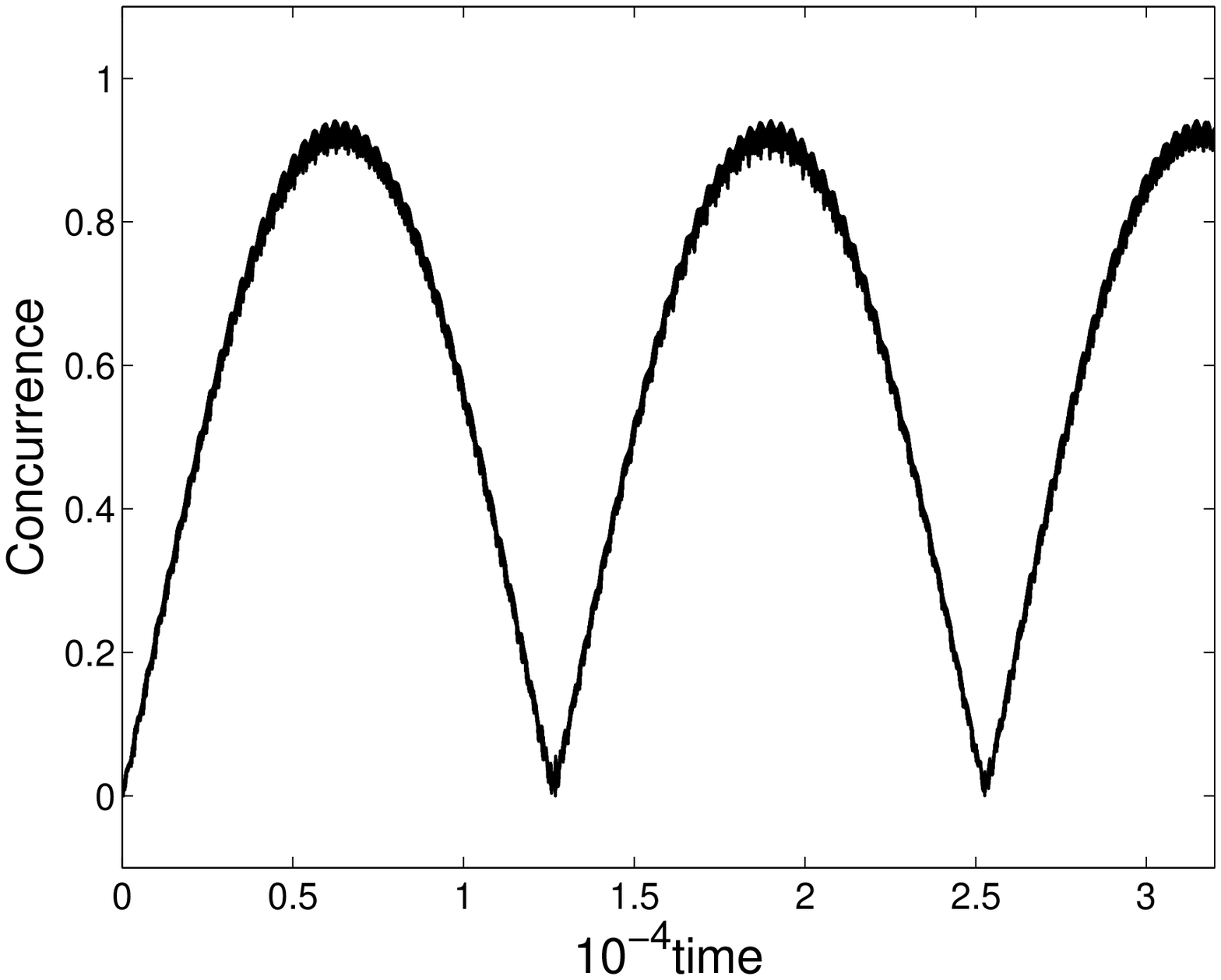}}
\subfigure[$|\psi(0)\rangle=|11\rangle$, triangular field]{
\label{figent11:tri}
\includegraphics[width=3in]{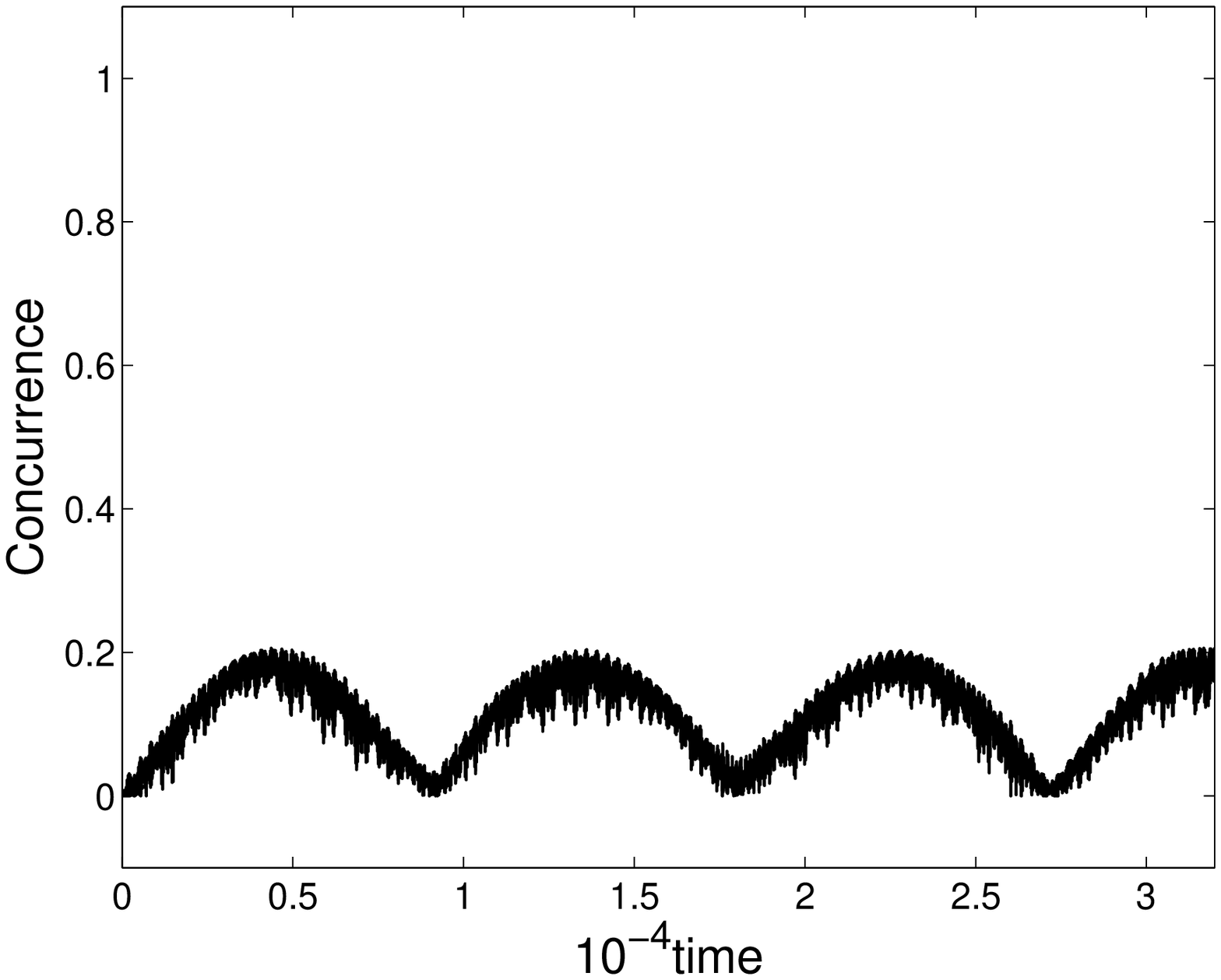}}
\subfigure[$|\psi(0)\rangle=|11\rangle$, no field $A=0$]{
\label{figent11:std}
\includegraphics[width=3in]{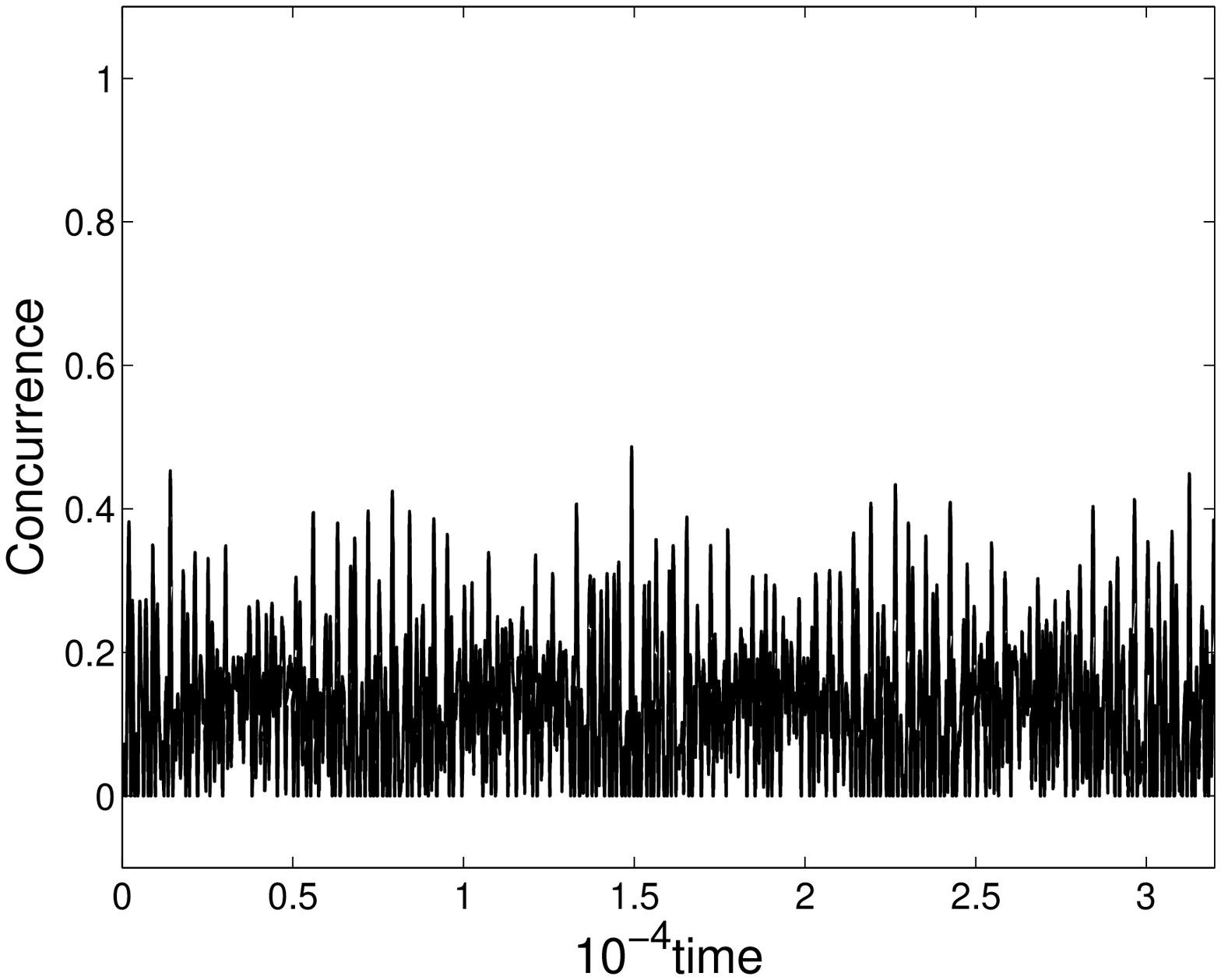}}
\caption{The time-evolution of entanglement for
$|\psi(0)\rangle=|11\rangle$ under different field vs. the pure
dissipation process without any modulating action}\label{figent11}
\end{figure}

The calculated entanglement evolution are plotted in the figures of
\ref{figent00}, \ref{figent01}, \ref{figent11} according to
different kinds of modulating field with different initial condition
for the two excitons in quantum dots $|\psi(0)\rangle$. Fig.
\ref{figent00} shows the concurrence for the initial pure state
$|\psi(0)\rangle= |00\rangle$ under rectangular wave
\ref{figent01:rec}, cosine wave \ref{figent01:cos}, triangular wave
\ref{figent01:tri} and without external field \ref{figent01:std}.
Obviously, the entanglement fluctuations in sub-figure
\ref{figent01:std} are greatly decreased by the periodical external
field and the entanglement evolution is improved to different
extent. The effect of cosine wave \ref{figent01:cos} is better than
the others. And Fig. \ref{figent01} and Fig. \ref{figent11} display
the concurrence for initial condition of $|\psi(0)\rangle=
|01\rangle$ and $|\psi(0)\rangle= |11\rangle$, respectively. It is
easy to see that $|01\rangle$ and $|00\rangle$ or $|11\rangle$ can
be distinguished by the concurrence evolution driven by the same
magnetic field. \\

Through the driving of the magnetic field, a much steady and high
entanglement between the two quantum dots can be prepared. It
results from the adjustment of the energy bias due to the field
$F(t)$. In some interval of one period of $F(t)$, the energy bias is
reduced, so that the probability of the exciton transition is
greatly increased due to the Possion probability distribution of the
multiple photon process \cite{Mandel, Scully}. Then the probability
that the two excitons make transition simultaneously is also much
greater than that without such a periodical field. When this kind of
process happens, the two excitons gain some spatial correlation. The
correlation helps to increase entanglement between them. And in
other interval of the same period, the energy bias is larger than
that without such a field, it could help to maintain the entangled
state just gained by the above correlation process. At the same
time, the spontaneous radiation happens randomly which reduces the
concurrence. \\

Three important features need to be noticed: (i) The effect of the
single-mode cavity, $g\sum_{i=1}^2(a+a^+)\sigma^x_i$ in equation
\ref{Hami}, is twofold. On the one hand, indirect interaction
between the two excitons is induced by cavity mode. If both excitons
could transfer synchronously, they are correlated at the same time,
which makes them further entangled in space. However that
probability is very small due to the condition $\omega\ll\epsilon$.
On the other hand, it is quantum noise \cite{Zoller} that effects on
the entangled pair. Therefore, there are great fluctuations in the
time region as in the last sub-figure of figures \ref{figent00},
\ref{figent01}, \ref{figent11}. (ii) There is evident difference
between the evolution results from initial state $|01\rangle$ and
$|00\rangle$ or $|11\rangle$. For the initial state of $|01\rangle$,
we find all the three kinds of periodical fields can create and
maintain a highly entangled exciton-pair: in figure
\ref{figent01:rec}, the peak value of the first wave envelope is
$0.971332$ and around the same peak, the fidelity of $C\geq0.5$
persists a sufficient long time $14792$; in figure
\ref{figent01:cos}, the first peak value is $0.971071$ and around
the same peak, the time interval of fidelity $C\geq0.5$ is as long
as $20389$; in figure \ref{figent01:tri}, the first peak value is
$0.997605$ and around the same peak, the time interval of $C\geq0.5$
is $10899$. This advantage of the case $\psi(0)=|01\rangle$ over the
other two cases stems from the effect of the single-mode cavity
during the exciton transition. When the system is prepared in
$|01\rangle$ or $|10\rangle$, one exciton could drop from the
excited level and simultaneously the other exciton could absorb the
photons just emitted by the former. This is a virtual process by
which the two excitons are correlated. But when the system is in
$|00\rangle$ or $|11\rangle$, this kind of synchronization resorts
to the photons provided by the single-mode. That is a true process
whose probability is much lower. (iii) The rectangle wave or
triangle wave essentially consists of infinite cosine waves of
multiplicate frequencies. Thus the combined effect of entanglement
enhancement from them is not as good as that from a
fundamental-frequency cosine wave due to their asynchronous
evolution (to compare sub-figure (b) with (a) or (c) in figure
\ref{figent00}, \ref{figent01}, \ref{figent11}). \\

\begin{figure}[htbp]\centering
\subfigure[$|\psi(0)\rangle=|01\rangle$, rectangular field]{
\label{figentr01:rec}
\includegraphics[width=3in]{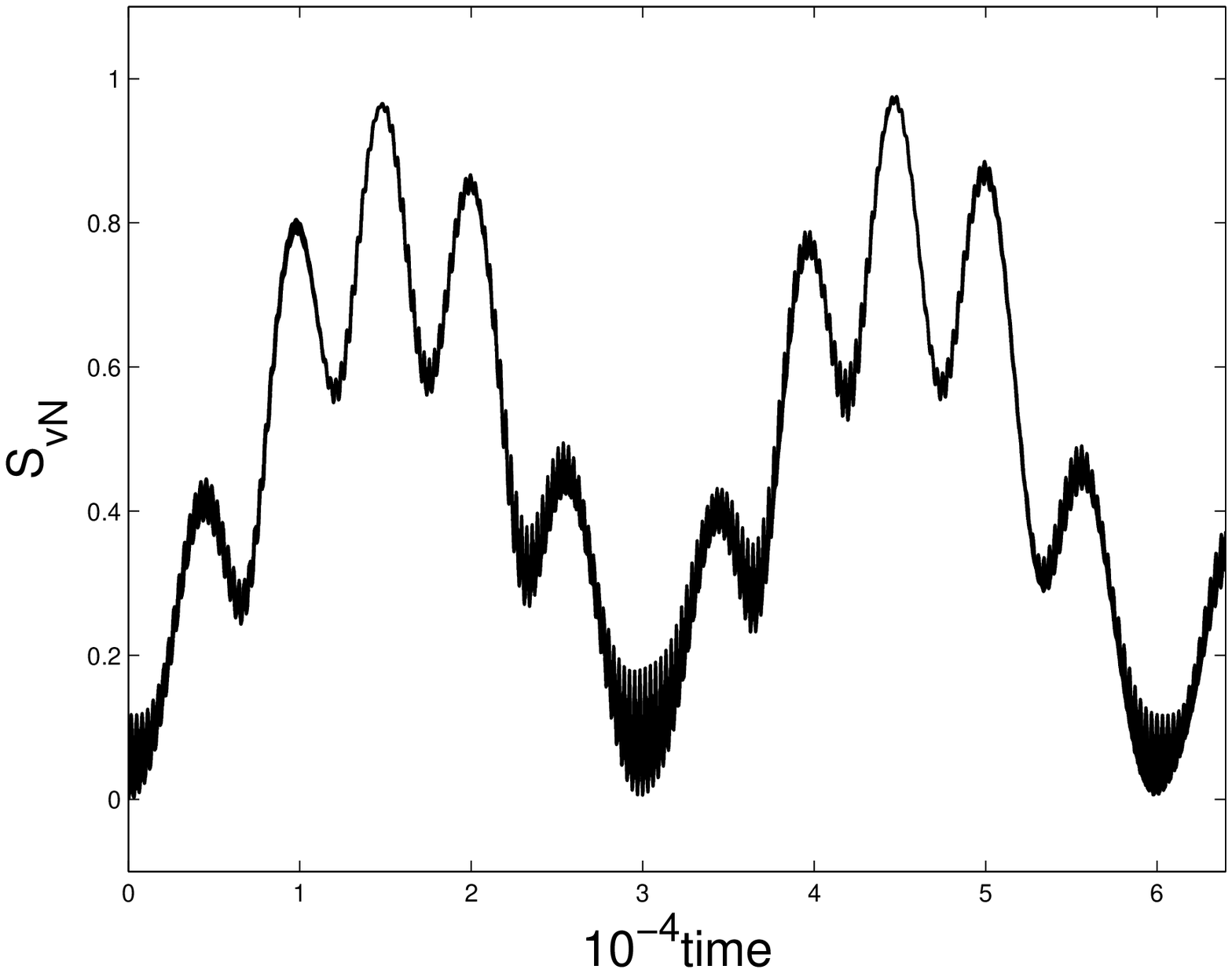}}
\subfigure[$|\psi(0)\rangle=|01\rangle$, cosine field]{
\label{figentr01:cos}
\includegraphics[width=3in]{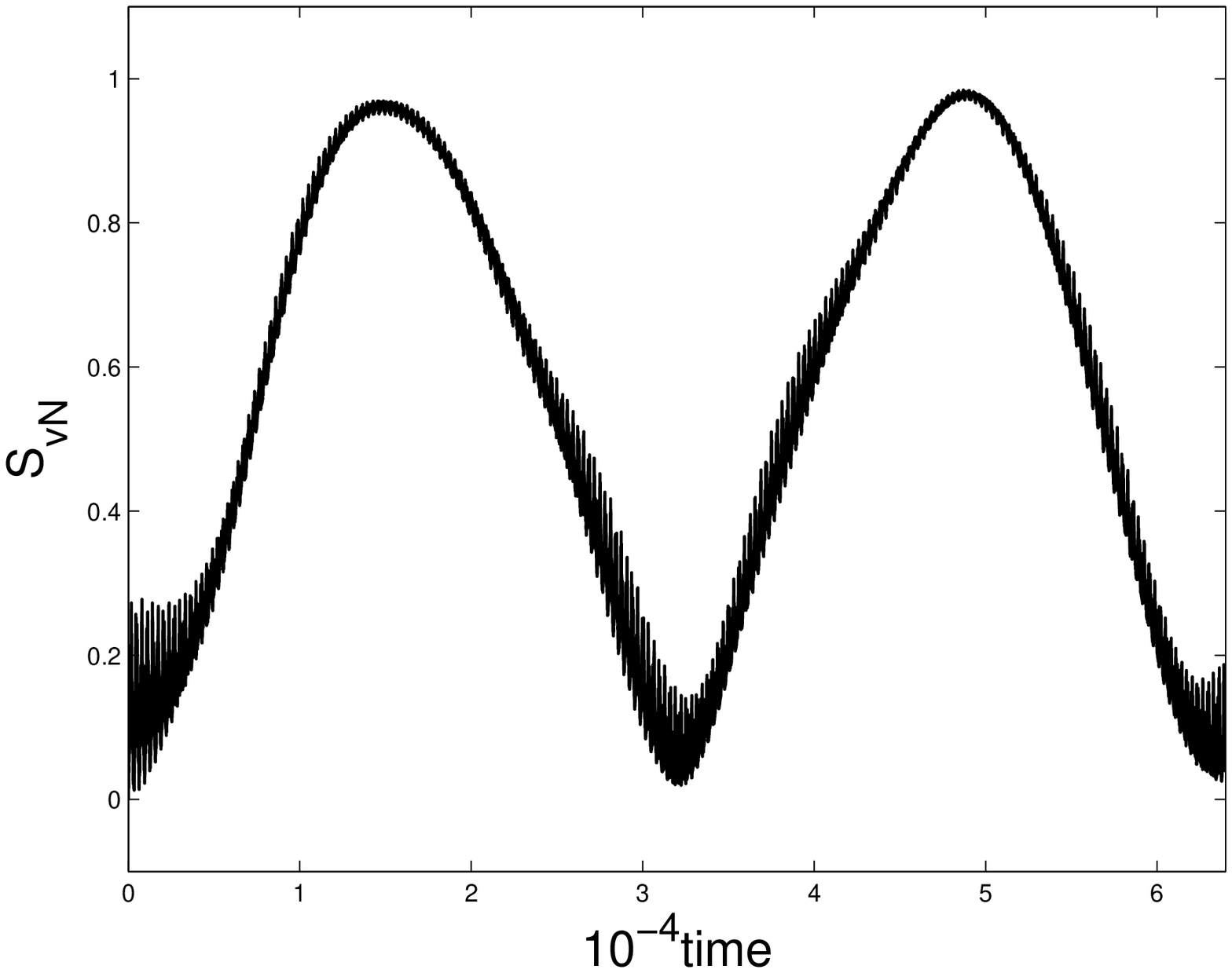}}
\subfigure[$|\psi(0)\rangle=|01\rangle$, triangular field]{
\label{figentr01:tri}
\includegraphics[width=3in]{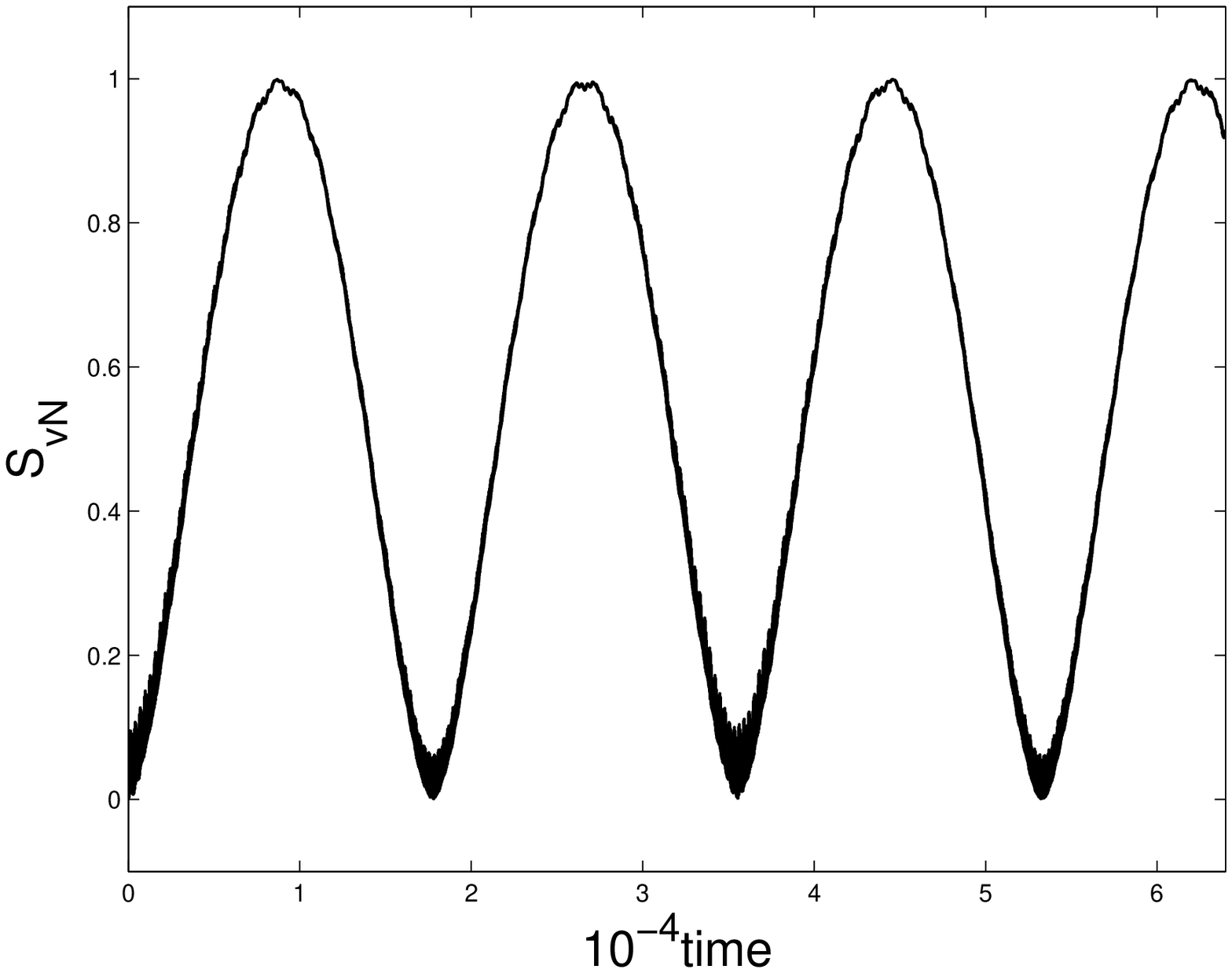}}
\subfigure[$|\psi(0)\rangle=|01\rangle$, no field $A=0$]{
\label{figentr01:std}
\includegraphics[width=3in]{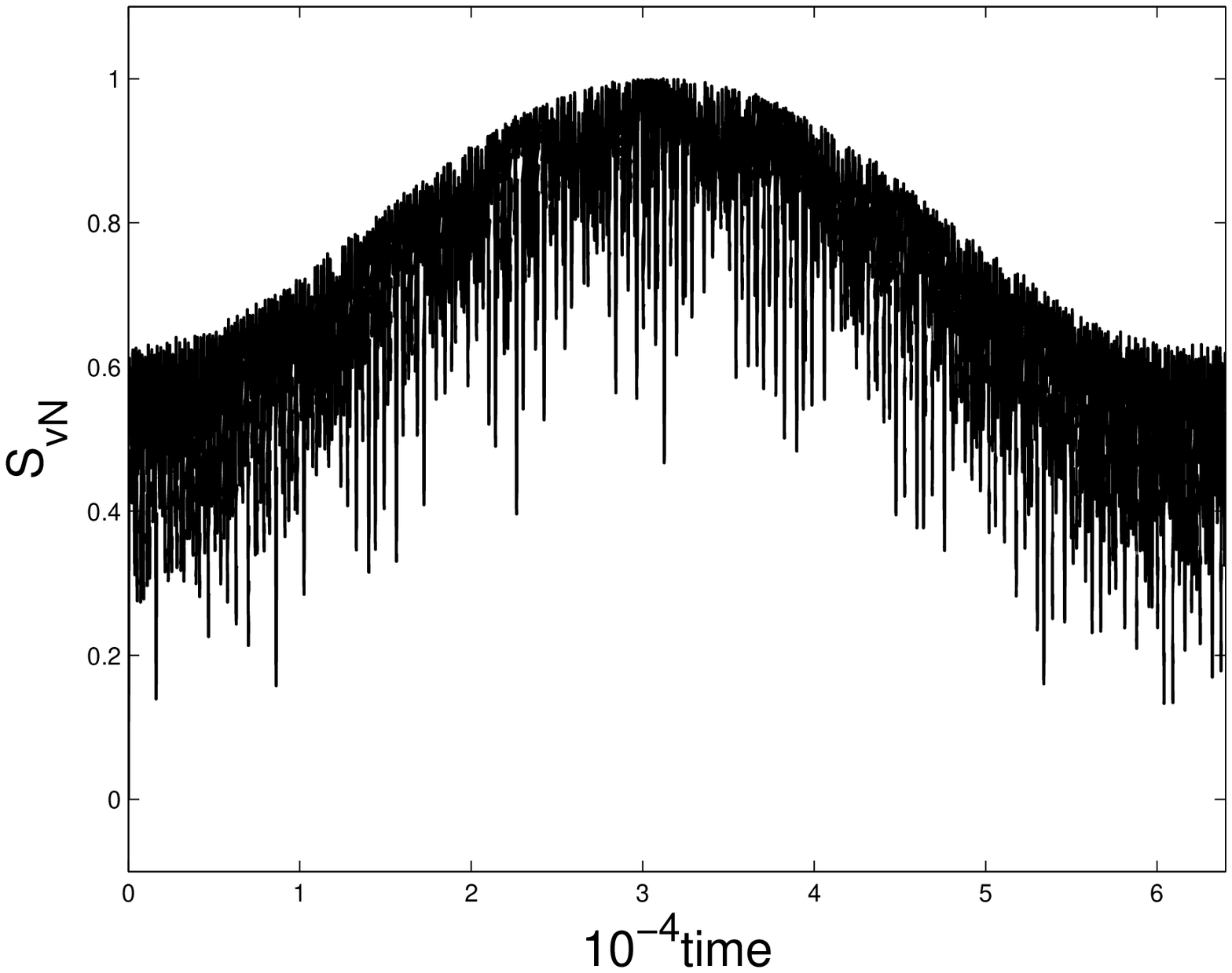}}
\caption{The time-evolution of von Neumann entropy for
$|\psi(0)\rangle=|01\rangle$ under different field vs. the pure
dissipation process without any modulating action}\label{figentr01}
\end{figure}

Since entanglement entropy can measure the amount of quantum
information inside an entangled pair of qubits, we give the time
evolution results of entropy for the initial pure state
$|\psi(0)\rangle= |01\rangle$ under rectangular wave
\ref{figentr01:rec}, cosine wave \ref{figentr01:cos}, triangular
wave \ref{figentr01:tri} and without external field
\ref{figentr01:std}. It is evident that the evolution behavior of
the von Neumann entropy is in good agreement with that of the
concurrence. It is further proved that highly-entangled pair of
qubits can be created by controlled external field.

\section{Conclusion}\label{conclusion}

In this paper, a entangled exciton pair is prepared in two separate
quantum dots embedded in a single-mode cavity by a periodical
external magnetic field. The two excitons are simplified as two
two-level atom or spins, which are initially prepared as a product
state, which means $C(\psi(0))=0$. The calculated results suggest
that it is possible to create steady highly-entangled pair by a
driving external field. And we find that an almost maximally
entangled state can be prepared from $\psi(0)=|01\rangle$ or
$\psi(0)=|10\rangle$. The results indicated that the cosine wave has
advantage over the rectangular and triangular field in entanglement
enhancement and persistence. Our approach can be used to study the
evolution of entanglement and quantum entropy for two quantum dots.
The revealed physical mechanism can be applied to control the
quantum devices in the future.

\end{document}